\documentstyle[12pt]{article}
\newcommand{\be}{\begin{equation}}
\newcommand{\ee}{\end{equation}}
\newcommand{\bea}{\begin{eqnarray}}
\newcommand{\eea}{\end{eqnarray}}
\newcommand{\al}{\alpha}
\newcommand{\bt}{\beta}
\newcommand{\gm}{\gamma}

\newcommand{\dl}{\delta}
\newcommand{\Dl}{\Delta}
\newcommand{\eps}{\epsilon}

\newcommand{\et}{\eta}
\newcommand{\th}{\theta}

\newcommand{\thv}{\vartheta}

\newcommand{\lm}{\lambda}
\newcommand{\ks}{\xi}
\newcommand{\rh}{\rho}
\newcommand{\rhv}{\varrho}
\newcommand{\sg}{\sigma}

\newcommand{\ups}{\upsilon}

\newcommand{\ch}{\chi}

\newcommand{\om}{\omega}
\newcommand{\Om}{\Omega}

\newcommand{\rarrow}{\rightarrow}
\newcommand{\Rarrow}{\Rightarrow}
\newcommand{\nn}{\nonumber}

\begin{document}
\title{Parametric resonant acceleration of particles 
by gravitational waves}
\author{K Kleidis, H Varvoglis and D Papadopoulos\\
{\small Section of Astrophysics, Astronomy and Mechanics}\\
{\small Department of Physics}\\
{\small Aristotle University of Thessaloniki}\\
{\small 54006 Thessaloniki, Greece} }

\maketitle
\begin{abstract}

We study the resonant interaction of charged particles with a gravitational wave 
propagating in the non-empty interstellar space in the presence of a uniform 
magnetic field. It is found that this interaction can be cast in the
form of a {\em parametric resonance} problem which, besides the main resonance, 
allows for the existence 
of many secondary ones. Each of them is associated with a non-zero resonant width, 
depending on the amplitude of the wave and the energy density of the interstellar 
plasma. Numerical estimates of the particles' energisation and the ensuing damping 
of the wave are given. 

\end{abstract}
\section{Introduction}

Despite the numerus efforts made up today to detect gravitational waves, there is no 
convincing evidence for their existence (Thorne 1987). This is due to the fact
that not only their amplitude is very small (Smarr 1979), but it is highly possible
that some kind of damping mechanism operates on them as they travel through space 
(Esposito 1971, Macedo and Nelson 1983, Papadopoulos and Esposito 1985). This damping
may originate in the interaction of the gravitational wave with the interstellar
matter (Macedo and Nelson 1990, Varvoglis and Papadopoulos 1992).

In a recent paper (Kleidis et al 1993, which hereafter is referred to as Paper I)
the problem of the interaction of a charged particle with a gravitational wave,
in the presence of a uniform magnetic field, has been modelled as a Hamiltonian 
dynamical system. The corresponding analysis was carried out for various
directions of propagation of the wave with respect to the magnetic field. It 
was found that, in the oblique propagation, diffusive acceleration of the particle, 
due to secular energy transfer from the wave, could lead to a damping.

The most important results, however, came out from the parallel propagation case 
where the dynamical system is trapped at an exact resonance between the Larmor 
frequency of the particles and the frequency of the wave. In this case a {\em phase 
lock} situation appears (e.g. see Menyuk et al 1987), leading to an "{\em infinite}" 
acceleration of the particle and, 
consequently, to a non-trivial damping of the wave. The zero probability problem 
of the exact resonance was waived out in a more recent paper (Kleidis et al 1995, 
which hereafter is referred to as Paper II), by considering that the propagation 
of the gravitational wave takes place in a space filled with plasma, which results 
in a dispersion of the wave (Grishchuk and Polnarev 
1980). In this case, resonant phenomena occur also in the quasiparallel case, i.e. 
propagation at a small angle with respect to the direction of the magnetic field, 
$\thv \leq 5^{\circ}$.

In the present paper we perform an elaborated study of the resonant 
interaction between charged particles 
and a linear polarized gravitational wave, which propagates in a non-empty 
space, parallel (or quasiparallel) to the direction of a uniform magnetic 
field ${\bf B}=B_{0}{\hat e_{z}}$. The interstellar plasma is represented by 
a collisionless gas of particles, where by collisionless we simply mean that 
the mean time between succesive collisions of the particles is much larger than 
the period of the gravitational wave. As we show, in this case, the generalized 
coordinate $x^1$ obeys 
a Mathieu differential equation (Abramowitz and Stegun 1970). As long as the 
resonance condition given by Eq. (23) of Paper II is fullfiled, this equation 
corresponds to a dynamical system which is trapped at a {\em parametric 
resonance}, where any external action ammounts to a time variation 
of the frequency parameter (Landau and Lifshitz 1976).  In this case, the system's 
equilibrium at rest $(x^1 = 0)$ is unstable. Any deviation from this state, 
however small, is sufficient to lead to a rapidly increasing displacement $x^1$. 
Then, the corresponding generalized momentum $\pi_1$ satisfies a modified Mathieu 
equation, resulting in an exponential increase of the perpendicular energy, $I_1$, 
of the particle. In this case, also, there exists a large number of secondary 
resonances of non-zero width (Bell 1957), at which the dynamical system may 
be trapped. Their exact location is given in terms of the energy parameter 
$I_0$. The fact that the total measure of the resonant widths is non-zero 
increases the effectiveness of the interaction mechanism under study, both in 
the process of the accelaration of particles and in the damping of the wave.

\section{Parametric resonance}

We consider the non-linear interaction between a charged particle and a linear 
polarized gravitational wave propagating in a non-empty space. The interaction 
takes place in the presence of a uniform and static, 
in time, magnetic field, ${\bf B}=B_{0}{\hat e_{z}}$, which does not interact with 
the gravitational wave. 
The motion of a charged particle in curved spacetime is given, in Hamiltonian 
formalism (Misner et al 1973) by the differential equations
\be
{d x^{\mu} \over d{\lm}} = {\partial H \over \partial {\pi}_{\mu}} \; , 
\; \; \; {d {\pi}_{\mu} \over d {\lm}} = -{\partial H \over \partial x^{\mu}}
\ee
where ${\pi}_{\mu}$ are the generalized momenta (corresponding to the 
coordinates $x^{\mu}$) and the "super Hamiltonian", H, is given by the relation
\be
H={1 \over 2} g^{\mu \nu} \left ( {\pi}_{\mu} - e A_{\mu} \right ) \left (
{\pi}_{\nu} - e A_{\nu} \right ) \equiv {1 \over 2}
\ee
(in a system of geometrical units where $\hbar = c = G = 1$). In Eq. (2) 
$g^{\mu \nu}$
denotes the components of the contravariant metric tensor, which are defined as
$$
g^{\mu \nu} = {\et}^{\mu \nu} + h^{\mu \nu}
$$
with ${\et}^{\mu \nu} = diag (1, -1, -1, -1)$ and $|h^{\mu \nu}| \ll 1$.
$A_{\mu}$ is the vector potential, corresponding to the tensor of the
electromagnetic field in a curved spacetime $F_{\mu \nu}$. The mass of the
particle is taken equal to 1. For the specific form of the magnetic field we
take
$$
A_{0} = A_{1} = A_{3} = 0 \; , \; \; \;  A_{2} = - B_{0} x^{1}
$$

According to Paper II the problem depends on four parameters: The normalized, dimensionless 
amplitude of the gravitational wave, $\al$, the angle of propagation with respect 
to the direction of the magnetic field, $\thv$, the dimensionless frequency, $\nu$, 
(which is the ratio between the frequency of the wave $ \om$ and the Larmor frequency 
of the particles $ \Om$) and the small parameter $ \bt \sim \rhv / \om^2 $, which is a  
dimensionless constant depending on the energy density of the interstellar plasma 
$ ( \bt \ll 1 ) $ (e.g. see Grishchuk and Polnarev 1980, Kleidis et al 1995).

Resonant phenomena become dominant, in particular, when the gravitational 
wave propagates parallel to the direction of the magnetic field, i.e. $\thv = 0^{\circ}$. 
In this case, the non-zero components of the metric tensor are
$$
g^{\mu \nu} = diag \left ( \;1, \; - {1 \over 1 - \al cos k_{\mu} x^{\mu}}, \; 
- {1 \over 1 + \al cos k_{\mu} x^{\mu}}, \; - 1 \right )
$$
and the super Hamiltonian, given by Eq. (2), is finally written in the form 
\bea
 H&=& {1 \over 2} ( 2 \nu - 1 ) \left ( 1 - {\nu -1 \over \nu} \bt \right ) I_0^2 \nn \\
  &-& f( \nu, \bt) I_0 I_3 - {1 \over 2} {\pi_1^2 \over  1 + \al sin \th^3 } - 
{1 \over 2} {(x^1)^2 \over 1 - \al sin \th^3}
\eea
where, for the sake of convenience, we have set
$$
f( \nu , \bt ) = \nu \left ( 1 - {2 \nu -1 \over 2 \nu} \bt \right )
$$
Since $\th^0$ is a cyclic coordinate, the corresponding generalized momentum $I_0$ 
is a constant of the motion and, according to Paper II, $I_0 \neq 0$.
Now, the equation of motion for $\th^3$ is readily solved to give
\be
{d \th^3 \over d \lm} = {\partial H \over \partial I_3} \; \; \; \; \; 
\Rarrow \; \; \; \; \; \th^3 = {\pi \over 2} - f( \nu , \bt) I_0 \lm
\ee
where $\pi / 2$ is the initial phase. Since $\al \ll 1$, in what follows we keep 
terms up to first order in $\al$. Using the approximation
$$
{1 \over 1 \pm \al sin \th^3} \approx 1 \mp \al sin \th^3
$$
the geodesic differential equation of motion for $x^1$ is written in the form
\be
{d^2 x^1 \over d \lm^2} + \al f( \nu , \bt )I_0 sin \left [ f( \nu , \bt )
I_0 \lm \right ]  {d x^1 \over d \lm } + x^1 = 0
\ee
where we have also used the equation of motion for $\pi_1$ and the fact that 
$$
\pi_1 = {d x^1 \over d \lm} + e A_1
$$
in which, for the specific form of the magnetic field, $A_1 = 0$. 

We consider a transformation of the independent variable, $\lm$, of the form 
$s = s( \lm )$, where $s (\lm )$ is chosen as to satisfy the 
differential equation
\be
{d^2 s \over d \lm^2} + \al f ( \nu , \bt ) I_0 sin \left [ f( \nu , \bt) I_0 \lm 
\right ] \; \; {d s \over d \lm} = 0
\ee
The condition (6) guarantees that the resulting, in terms of $s$, equation of motion 
for $x^1$ will continue to 
represent the differential equation of a geodesic (Papapetrou 1974). Eq. (6) 
may be solved in the first order to $\al$, to give
\be
s ( \lm ) \simeq \Om \left ( \lm + {\al \over f( \nu , \bt ) I_0} sin [ 
f (\nu , \bt) I_0 \lm ] \right ) 
\ee
where the Larmor frequency, $ \Om $, appears as an integration constant. 
Now, in terms of $s(\lm)$, Eq. (5) may be written in the form
\be
{d^2 x^1 \over d s^2} \; \; \;  + \; \; \; \Om^2 \left ( 1 - 2 \al \; cos [ f( 
\nu , \bt ) I_0 \Om s ] \right ) x^1 \; \; = \; \; 0
\ee
where we have also used the fact that $ \al \ll 1$. 
This is a Mathieu equation (Abramowitz and Stegun 1970) which is closely related 
to the problem of {\em parametric resonance} in dynamical systems (Landau and 
Lifshitz 1976). According to it, the external action (in our case the 
gravitational wave) amounts only to a periodic time variation of the frequency 
parameter of the unperturbed system (the Larmor frequency, $\Om$). Indeed, in 
this case, the generalized frequency 
\be
\sg^2 (s) = \Om^2 [ 1 - 2 \al \; \; cos ( \gm s)]
\ee
differs only slightly from the constant $\Om$ and is a simple periodic function.  In
Eq. (9) $ \gm = f( \nu , \bt) I_0 \Om $ denotes the frequency of this periodic function. 
It can be proved (Landau and Lifshitz 1976) that the main parametric resonance in a 
dynamical system appears when $ \gm = 2 \Om $.

It has been shown (Kleidis et al 1995) that the system gravitational wave $+$ 
charged particle is trapped in a resonance when the resonant condition, given by Eq. (23) 
of Paper II, is satisfied, that is when
\be
f ( \nu , \bt ) I_0 = 2
\ee
which gives $ \gm = 2 \Om $. Therefore, once the condition (10) is satisfied, our 
dynamical system is actually trapped in a {\em main parametric resonance}. 
In general, there is a region of instability of non-zero measure, the so called 
{\em resonant width} (Landau and Lifshitz 1976), around the exact value $ \gm 
= 2 \Om $, inside which the parametric resonance mechanism operates, which is
\be
2 \Om - {1 \over 2} \Dl \eps \; \leq \; \gm \; \leq \; 2 \Om + {1 \over 2} \Dl \eps
\ee
The range (Bell 1957) of the resonant width is given, in our case, by $ \Dl 
\eps = 2 \al \Om $. 

Apart from the case where $\gm $ is close to $ 2 \Om $, parametric resonance 
also occurs when the frequency of the parameter $ \sg (s) $ 
is close to any value $2 \Om /n$, where $n$ is a natural number (Landau and 
Lifshitz 1976).
Therefore, except from the primary resonance, there is also a large number 
of secondary ones at which the dynamical system under consideration may be 
trapped. Each of them also has a non-zero resonant width, the exact form of 
which, in terms of $ \Om $, is given by (Bell 1957)
\be
\Dl \eps^{(n)} = ({n \over 2})^{2n-3} \al^n {1 \over [(n-1)!]^2} \Om
\ee
that is, it decreases rapidly with increasing $n$, as $ \al^n $.

In our case it is more convenient to express both the possition of each 
of the secondary parametric resonances and the corresponding resonant widths 
in terms of the constant of the motion $I_0$, which corresponds to a measure 
of the total energy of the dynamical system. Since, according to Paper II, the 
exact resonance is achieved for 
$$
I_0 = I_0^{\star} = {2 \over f(\nu , \bt)}
$$
the dynamical system will be trapped at a secondary one, of order $n$, when
\be
I_0 = {I_0^{\star} \over n} 
\ee
For each of these secondary resonances the corresponding resonant width 
is given by
$$
\Dl I_0^{(n)} = {\Dl \eps^{(n)} \over f ( \nu , \bt)} \; \; {1 \over \Om}
$$
or else
\be
\Dl I_0^{(n)} = {1 \over f} \; \; ({n \over 2})^{2n-3} \; \; \al^n \;  
{1 \over [(n-1)!]^2} 
\ee
where $f = f( \nu , \bt ) $. Using Stirling's formula, Eq. (14) is simplified 
to
\be
\Dl I_0^{(n)} = {4 \over \pi f} \; \; {1 \over n^2} \; \; \left ( {e \over 2} 
\sqrt { \al } \right )^{2n}
\ee
If, furthermore, we set
$$
\ch = {e^2 \over 4} \al \;  < \;  1 
$$
the resonant width of a parametric resonance of order $n$ is finally written in the 
form
\be
\Dl I_0^{(n)} = {4 \over \pi f} \; \; {\ch^n \over n^2} 
\ee
In this case, we may, also, calculate the {\em total measure} of the resonant 
widths, in terms of $ I_0 $, which corresponds to the probability of 
the dynamical system to be trapped in a resonance. The total measure of 
the resonant widths is written in the form
\be
\sum \Dl I_0 \; \; = \; \; {4 \over \pi f} \; \sum_{n=1}^{\infty} {\ch^n 
\over n^2}
\ee
As long as $ \ch < 1 $, the series on the r.h.s of Eq. (17) satisfies D' Alembert's 
criterion (e.g. see Gradshteyn and Ryzhik 1965) and therefore converges absolutely 
and uniformly. In this case the total measure of the resonant widths can be 
written as
\be
\sum \Dl I_0 \; \; = \; \; {4 \over \pi f} \; \ch \; _3F_2 (1, \; 1, \; 
1, \; ; 2, \; 2 \; ; \ch )
\ee
where we have taken into account the properties and the series expansion of the 
generalized hypergeometric series $ _3F_2 (1, \: 1, \: 1; \: 2, \: 2 ; \: \ch 
) $ (e.g. see Erdelyi et al 1953). If we keep only terms of first order with respect to 
$\al $ and $ \bt $, Eq. (18) is written in the form
\be
\sum \Dl I_0 \; = \; {e^2 \over \pi \nu} \; \al \; \; + \; \; O ( \al \bt ) 
\; \; + \; \; O ( \al^2 )
\ee
where $e$ is the basis of the natural logarithms.

\section{Numerical results}

The most interesting case of interaction between 
a gravitational wave propagating in a non-empty space and a charged particle 
is the quasiparallel case, in which the propagation takes place 
at a small angle with respect to the direction of the magnetic field. 

Since, in this case, the phase velocity of the gravitational wave is greater 
than the velocity of light (Grishchuk and Polnarev 1980), it is probable that 
its projection along the $ z $-axis, $ \ups_z = \ups_{ph} 
cos \thv $, to be exactly equal to $c$. This leads (see Paper II) to a combined 
interaction (both chaotic and resonant) between the particle and the wave, 
resulting in an overall decrease in the parallel momentum 
$ I_3 $ (which corresponds to an overall increase of the perpendicular one $I_1$) 
as a function of $ s $ (see Fig. 1a). 
\\
\\
%\mbox{%\epsfxsize=8.cm %\epsfysize=7.cm %\epsfbox{fig1a.eps} %\epsfxsize=8.cm 
%\epsfysize=7.cm %\epsfbox{fig1b.eps}} \\
{\small {\bf Fig. 1a and b:} Plots of $I_3$ versus $s$ (the affine parameter) 
for a numerically integrated trajectory with $\thv = 0.5^{\circ} $, $\nu = 52.01$, 
$ \bt = 10^{-5} $ and $\al = 0.002$. Notice the overall decrease in $I_3$ 
when {\bf (a)} $ I_0 = I_0^{\star} = 0.038 $, in contrast to the 
purely chaotic 
behaviour when {\bf (b)} $I_0 = 0.0385 $ which lies outside the corresponding 
theoretical resonant width $\Dl I_0 = 7 \cdot 10^{-5} $.}
\\

Since our problem corresponds to a parametric resonance, except 
from the primary resonant condition, for $I_0 = I_0^{\star} $, there is 
also a large number of values of the constant of motion $ I_0 $, $ I_0 = 
I_0^{\star} / n \; (n = 1, \: 2, \: 3, ...) $, leading to secondary resonances, 
each of which has a non-zero resonant width. One expects that the above 
mentioned theoretical results in the quasiparallel propagation case 
should continue to hold around each of these secondary resonances, 
provided that the values of $I_0$ lie within the corresponding 
resonant width, $ \Dl I_0^{(n)} $ (in connection to this see Fig. 1b). 

To check for the validity of these theoretical estimates,
we integrate numerically the equations of motion of a charged particle 
in a gravitational wave for several values of the constant of motion $ I_0 $, when 
$I_0^{\star} = 0.038$. Some of the corresponding results are presented in Figs. 
2 and 3. According to these, when $I_0 = I_0^{\star} / n $, we do observe the 
expected decrease in $ I_3 $, which corresponds to the exponentially increasing 
brantch of $I_1$, and implies that our dynamical system is trapped in a 
parametric resonance.
\\
\\
%\mbox{%\epsfxsize=8.cm %\epsfysize=7.cm %\epsfbox{fig2a.eps} %\epsfxsize=8.cm 
%\epsfysize=7.cm %\epsfbox{fig2b.eps}} \\
{\small {\bf Fig. 2a and b:} Plots of $I_3$ versus $s$  
for a numerically integrated trajectory with $\thv = 0.5^{\circ} $, $\nu = 52.01$, 
$ \bt = 10^{-5} $ and $\al = 0.002$ when {\bf (a)} $ I_0 = I_0^{\star} / 2 $ 
and {\bf (b)} $ I_0 = I_0^{\star} / 5 $}
\\
\\
The fact that the interaction of a charged particle with a gravitational 
wave in the parallel and/or the quasiparallel case corresponds to a parametric 
resonance, may impose noteworthy implications to the problem of the interaction 
between charged particles and a gravitational wave, increasing both the efficiency 
and the probability to achieve this particular interaction mechanism.
\\
%\mbox{%\epsfxsize=8.cm %\epsfysize=7.cm %\epsfbox{fig3a.eps} %\epsfxsize=8.cm 
%\epsfysize=7.cm %\epsfbox{fig3b.eps}} \\
{\small {\bf Fig. 3a and b:} Similar to Figs. 2, except that {\bf (a)} 
$ I_0 = I_0^{\star} / 9 $ and {\bf (b)} $ I_0 = I_0^{\star} / 12 $} 
\\
\\
As a gravitational wave propagates 
in the non-empty interstellar or intergalactic space, it encounters several
clouds with various characteristic temperatures and densities, while at the 
same time, it intersects 
the cosmic magnetic fields at various angles (Hillas 1984). In each 
case corresponds a different set of values of $ \bt $, $ \nu $ and $ \thv $. 
One expects that, at least for some of these sets, one of the resonant 
conditions is satisfied (since the corresponding resonant widths 
are non-zero), leading to a resonant interaction between a charged 
particle and the gravitational wave. 
According to Paper II, the result of this interaction is the acceleration of the 
particle at very high energies. Since the total energy $I_0$ is constant, the 
dynamical system (gravitational wave and charged particle) is {\em isolated}. 
Therefore any energisation of the particle corresponds to a damping of 
the wave.

It is worth to note that the decrease of $I_3$ in Figs. 2 and 3 is more pronounced 
for large values of $ n $. This is in contrast to what one may have expected, since 
the higher order resonances are considered to be weaker than the lower order ones. 
However this result 
is only phenomenological and of no physical importance, since it has to do only 
with the slope of the straight line $ I_3 \: - \: I_1 $. 
In complete correspondence with Paper II we may find, from the condition $ H_0 \simeq 1 / 2 $, 
that, in momentum space, the particles move along the straight line 
\be
I_3 \; = \; - \: {1 \over 2} \left ( {I_0^{\star} \over I_0} \right ) \: I_1 \; 
+ \; \left [ {2 \nu - 1 \over \nu^2} (1 + \bt ) {I_0 f( \nu , \bt ) \over 
2} \: - \: { 1 \over 4 } \left ( {I_0^{\star} \over I_0} \right ) \: \right ]
\ee
and since, at every $ n^{th}$-order parametric resonance we have $I_0 = I_0^{\star} 
/ n $, Eq. (20) can be written in the form
\be
I_3 \; = \; - \: {n \over 2} \: I_1 \; + \; \left [ {1 \over n} \: { 2 \nu - 1 \over 
\nu^2} (1 + \bt ) \: - \: {n \over 4} \: \right ]
\ee
We note that for $ n = 1 $ (the primary resonance) Eq. (21) is reduced to the 
corresponding one of Paper II [Eq. (28)]. In the present case, as $ n \rarrow 
\infty $, the straight line $ I_3 (I_1) $ tends to become parallel to the 
$I_3$-axis. This simply means that a small variation in $I_1$ leads to a large 
variation of $I_3$, which is exactly the case shown in Figs. 2 and 3.

However, there is another numerical result which may be of physical 
importance. It is related to the fact that, for several of the secondary 
resonances, resonant phenomena are observed even for values of $ I_0 $ lying 
outside 
of the corresponding, theoretically estimated, resonant widths [Eq. (14)]. 
It should be noted that this purely numerical result is not observed in the 
case of the 
primary resonance, $ I_0 = I_0^{\star} $. Some of these results are shown in 
Figs. 4. 

In Fig. 4a we present a plot of $I_3 $ versus $s$ for a numerically 
integrated trajectory with $I_0 = {I_0^{\star} \over 2} + 10^{-4} $. We see 
that an overall decrease in $ I_3 $ occurs although, in this case, $ \Dl 
I_0^{(2)} \simeq 10^{-7} $. The same is also true for the case presented in Fig. 4b, 
where $ I_0 = {I_0^{\star} \over 9} + 10^{-6} $, far away from the corresponding 
resonant width, which, in this case is $ \Dl I_0^{(9)} \simeq 10^{-27} $. 
Therefore, it seems that the numerically estimated resonant 
width, $ \dl I_0^{(n)} $, around each of the higher-order parametric resonances 
are many orders of magnitude larger than the corresponding theoretical one.
\\
\\
%\mbox{%\epsfxsize=8.cm %\epsfysize=7.cm %\epsfbox{fig4a.eps} %\epsfxsize=8.cm 
%\epsfysize=7.cm %\epsfbox{fig4b.eps}} \\
{\small {\bf Fig. 4a and b:} Plots of $ I_3 $ versus $s$ for a numerically 
integrated trajectory corresponding to a problem of parametric resonance 
which is {\bf (a)} similar to the one of Fig. 2a, but with $ I_0 = 0.0191 
= {I_0^{\star} \over 2} + \dl I_0^{(2)} $ and {\bf (b)} similar to the one of 
Fig. 3a, but with $ I_0 = 0.004221 = {I_0^{\star} \over 9} + 
\dl I_0^{(9)} $. In any of these cases $ \dl I_0^{(n)} \gg \Dl I_0^{(n)} $, 
but also $ \dl I_0^{(n)} \ll \sum \Dl I_0 $.}
\\
\\
\\
This behaviour results probably from the fact that, because of the small value 
of $I_0^{\star}$, the secondary resonances are closely spaced. Therefore, 
even a small diffusion process due to the chaotic motion, which is always present 
in this case, could probably bring the value of $I_0$ 
into a resonance, even if, initially, it was outside of it. 
Indeed, as we may see from Figs. 4, the resonant behaviour becomes evident 
only after a considerable time interval. This time interval should correspond 
to the correlation time of the diffusive process (Farina et al 1993). 
Therefore, our system may be a lot more efficient, with respect to the resonant 
interaction procedure, than the corresponding theoretical one. Notice also, 
that these phenomena are not present in the case of the primary resonance, 
since the corresponding value of $I_0$ lies far away from every of the 
secondary ones and hence, it possesses a well defined resonant 
width. For this reason, in this case, theoretical and numerical results are 
similar.

\section{Exact solutions}

In Paper II it has been shown that, when the gravitational wave propagates 
exactly parallel to the direction of the magnetic field (the $z$-axis), we 
may solve the equations of motion of the charged particle. Their solution 
reveals a {\em phase-lock} situation (Menyuk et al 1987, Karimabadi et al 
1990) resulting in an "{\em infinite}" acceleration of the particle along the 
$x$-axis. However, there, the solution was obtained through 
an averaging technique (e.g. see Lichtenberg and Lieberman 1983).
Since the problem is identified as that of a parametric resonance, 
in which the equation of motion along the $x$-axis corresponds to a Mathieu 
equation (Gradshteyn and Ryzhik 1965, Abramowitz and Stegun 1970), we may find 
exact solutions without resorting to any averaging technique.

In the parallel propagation case, $\thv = 0^{\circ} $, the dynamical system is 
trapped in a resonance when the condition (10) is satisfied. In this case the 
equation of motion (8), for the generalized coordinate $x^1$, is written in the 
form
\be
{d^2 x^1 \over d s^2} \; + \; \Om^2 [ 1 - 2 \al \: cos ( 2 \Om s ) ] \: x^1 \; 
= \; 0
\ee
or else 
\be
{d^2 x^1 \over d \ks^2} \; + \; \Om^2 [ 1 - 2 \al \: cos ( 2 \ks ) ] \: x^1 \; 
= \; 0
\ee
where we have set $\ks = \Om s $. This is a Mathieu equation with $a = 1$ and 
$q = \al$ (Gradshteyn and Ryzhik 1965, Abramowitz and Stegun 1970). Its solution 
is therefore given in terms of the Mathieu functions $ ce_1r $ and $ se_1$, 
associated with even and odd periodic solutions respectively. Each one of these 
solutions has a period of $\pi$ or $2 \pi$. 
In the present paper we consider periodic solutions of period $2 \pi$. It can be 
shown that, for a given point $(a, q)$ in the parameter space, there can be at 
most one periodic solution 
of period $ 2 \pi $ (Abramowitz and Stegun 1970) and therefore our solution is 
unique. The general solution of Eq. (23) is written in the form
\be
x^1 ( \ks ) = A \: ce_1 (\ks , \al ) + B \: se_1 ( \ks , \al )
\ee
where $A, \: B$ are constants.

For $\al \ll 1$ the Mathieu functions $ce_1$ and $se_1$ may be expanded in power 
series of $ \al $. Considering that $A = C = B$, in the linear approximation 
to $\al$ (e.g. see Abramowitz and Stegun 1970, Eq. 20.2.27, p. 725), Eq. (24) 
becomes
\be
x^1 = \sqrt 2 C sin \left ( {\pi \over 4} + \ks \right ) - \al { \sqrt 2 \over 8} 
C sin \left ({\pi \over 4} + 3 \ks \right )
\ee
In this case, using the normalization conditions of the Mathieu functions (e.g. see 
Gradshteyn and Ryzhik 1965, Eqs. 6.911.3, 6.911.5. and 6.911.7), we may determine 
the exact value of the constant $C$. The {\em mean square displacement} of 
$x^1$ is defined as
\be
< \: (x^1)^2 \: > = {1 \over 2 \pi} \: \int_{0}^{2 \pi } \left [ x^1( \ks ) \right ]^2 
d \ks = C^2
\ee
so that 
\be
C = \sqrt {< \: (x^1)^2 \: >}
\ee
Now, Eq. (25) is written in the form
\be
x^1 (s) = \sqrt 2 \: \sqrt {< \: (x^1)^2 \: >} sin \left ( {\pi \over 4} + 
\Om s \right ) - \al { \sqrt 2 \over 8} \sqrt {< \: (x^1)^2 \: >} sin \left 
( {\pi \over 4} + 3 \Om s \right )
\ee
and therefore, for $ \al = 0 $, initially $ (s = 0) $ we have 
\be
x_0^1 \: (s = 0) \; = \; \sqrt {< \: (x^1)^2 \: >}
\ee
corresponding to a sort of a Brownian motion (a not unexpected result).

However, the most important results come out from the solution of the 
equation of motion for the generalized momentum $\pi_1$
\be
{d \pi_1 \over d \lm} = {x^1 \over  1 - \al sin \th^3 }
\ee
To solve this equation we follow the same procedure as in the case of the 
generalized coordinate $x^1$. We then find that the perpendicular momentum 
$ \pi_1 $ obeys a modified Mathieu equation (e.g. see Abramowitz and Stegun 
1970), of the form
\be
{d^2 \pi_1 \over d s^2} \; - \; \Om^2 \: [ 1 - 2 \al cosh (2 \Om s) ] \: \pi_1 = 0
\ee
Its general solution is given in terms of the {\em radial} Mathieu functions 
$Ce_1$ and $Se_1$ (Gradshteyn and Ryzhik 1965, Abramowitz and Stegun 1970)
$$
\pi_1 (s) =  A \: Ce_1 ( \Om s, \al ) + B \: Se_1 ( \Om s , \al ) 
  =  A \: ce_1 (i \Om s , \al) - i B \: se_1 (i \Om s , \al ) 
$$
which, in the first approximation to $\al$, is written in the form
\be
 \pi_1 =  A \left [ cosh ( \Om s ) - { \al \over 8} cosh (3 \Om s ) \right ] 
 + B \left [ sinh ( \Om s ) - { \al \over 8} sinh (3 \Om s ) \right ]
\ee
Compatibility of these results with those of Paper II requires that, in this case, 
we have to set $ B = 0 $ thus obtaining
\be
\pi_1 = A \left [ cosh ( \Om s ) - {\al \over 8} cosh ( 3 \Om s) \right ]
\ee
From Eq. (33) it is evident that there is an exponential increase in the 
perpendicular momentum of the particles, which, in the zeroth order 
approximation to $\al$, corresponds to the exponential brantch of Eq. (27) of 
Paper II. We also note that, in the exact solution case, an additional, secularly 
increasing term, $ \sim \al \: cosh (3 \Om \lm) $, arises. 
The corresponding perpendicular energy, $I_1( \lm)$, in the first order approximation to $\al$, 
is given by
\be
I_1 \sim {1 \over 2} cos^2 \left ( \lm + {\al \over 2} sin 2 \lm \right ) 
\: cosh ( \al \lm ) \: \: \left [ 1 + {\al \over 2} \: cos ( 3 \lm ) \: 
cosh (3 \al \lm) \right ] +...
\ee
where we have used Eq. (7). Eq. (34) is in complete agreement to the 
results of Papers I and II. We see 
that the exponential increase in the perpendicular energy is, furthermore, 
modulated by periodic functions of time, which were smoothed out by the averaging 
technique in the previous Papers. It is important to point out that the 
phase-lock situation breaks down when $I_1$ becomes large enough, since then 
the perturbation term of the Hamiltonian, $H_1$, becomes comparable to $H_0$ 
and our approximation, in terms of the small parameter $\al$, is not valid 
any more.

Based on these results we may give an estimate of the absorption power per unit 
area of the gravitational wave, as a function of the proper distance from its 
source to Earth, for typical parameters of the interstellar gas.

In physical units, the energy density gained by the charged particles due to their 
resonant interaction with an incident gravitational wave, within a proper time 
interval $\Dl \lm$, is 
\be
\Dl E_{gained} = \Dl I_1 m_P c^2 \left ( {n_{act} \over n_{tot}} \right ) n_{tot}
\ee
where $m_P$ is the proton's mass, $(n_{act} / n_{tot})$ is the ratio of particles 
which interact resonantly with the gravitational wave, $n_{tot}$ is the particles' 
number density of the interstellar matter and $\Dl I_1$ is the energy change within 
the interval $\Dl \lm$, measured in dimensionless units.

Using Eq. (29) of Paper II, also translated in physical units, for $\nu = 1$ 
we obtain
\be
\Dl E_{gained} = {1 \over 2} \Dl I_1 m_P c^2 \sqrt { {m_P c^2 \over 2 \pi 
k_B T}} \bt n_{tot}
\ee
Accordingly, we consider that the interstellar space is filled with a 
non-relativistic perfect proton gas (Polnarev 1972), for which
\be
\bt = 16 \pi G {\rh \over \om^2} {k_B T \over m_P c^2} \; \; , \; \; 
\rh = m_P n_{tot}
\ee

We take $n_{tot} = 1 particle/cm^3$ and $T = 10 K$ as typical mean values for 
the interstellar space. Then, we find that the total energy density gained by 
the charged particles due to their resonant interaction with the gravitational 
wave, within the proper time interval $\Dl \lm$, is
\be
\Dl E_{gained} = 1.61 \times 10^{-39} {\Dl I_1 \over \om^2} \; \; \; 
({erg \over cm^3})
\ee
Since, on the other hand, the system wave $+$ particles is {\em isolated}, Eq. (38) 
corresponds also to the total energy density lost from the gravitational wave, due 
to the interaction mechanism under consideration.

The corresponding gravitational energy flux (power per unit area) lost in the same 
proper time interval is
\be
\Dl F_{lost} = 1.21 \times 10^{-29} {\Dl I_1 \over \om^2} \; \; \; 
({erg \over cm^2 sec})
\ee
Using Eq. (27) of Paper II for $\Dl I_1$, we integrate Eq. (39) over $\lm$, to obtain 
the total loss of gravitational energy flux during the whole proper time interval 
$0 \leq \lm \leq \lm_{E}$, at which the wave travels from its source to Earth. 
Taking into account that for every realistic proper time interval $\lm_{E}$ we 
have $\al \lm_{E} \ll 1$, we obtain
\be
F_{lost} = 6.1 \times 10^{-30} {1 \over \om^2} \al^2 \lm_E^2
\ee
where $\lm_E$ is normalized to ${c \over \Om}$. In physical units, 
Eq. (40) is written in the form
\be
F_{lost} = 6.1 \times {1 \over \om^2} \al^2 ({c \over \Om})^2 \lm_E^2
\ee
Now, $\lm_E$ is measured in $sec$ while $\om$ and $\Om$ in $Hz$. However, 
since $\nu = 1$ we have $\om = \Om$. Therefore, Eq. (41), in terms of the 
proper distance $r_E$ between the source of the gravitational waves and the 
Earth, is finally written in the form
\be
F_{lost} = 6.1 \times 10^{-30} {1 \over \om^4} \al^2 r_E^2
\ee
where $r_E$ is measured in $cm$.

The total gravitational energy flux expected in the neighbourhood of the Earth, 
if we do not take into account any interaction with the interstellar matter 
(Misner et al 1973, Thorne 1987), is
\be
F_{expected} = {c^3 \over 16 \pi G} \om^2 \al^2 = 8.1 \times 10^{36} \om^2 \al^2
\ee
Dividing by parts Eqs. (42) and (43) we obtain
\be
{F_{lost} \over F_{expected}} = 0.753 \times 10^{-66} \left ({Hz \over \om} \right )^6 
\left ({r_E \over cm} \right )^2
\ee
Eq. (44) gives the percentage of the flux lost due to the resonant interaction between 
a gravitational wave and a distribution of charged particles, at a proper distance 
$r_E$ from the source, over the theoretically expected gravitational energy flux, 
at the same distance, in the absence of resonant damping mechanisms.

It would be interesting to give some numerical examples of the flux damping, for 
realistic sources of gravitational waves in our galaxy. As those we choose the 
gravitational systems of bright NS/NS binaries (Thorne 1995). The gravitational 
waves generated by these sources have very low frequencies, $f=10^{-4} \; Hz$. 
Since $\nu = 1$ we have $\Om = 0.63 \times 10^{-3} \; Hz$ for protons, which is 
true for a magnetic field of strength $B \simeq 0.5 \times 10^{-6}$ G, a quite 
reasonable value for the interstellar space (Hillas 1984).

For gravitational waves generated from a binary NS/NS source at a distance 
$r_E = 10$ Kpc from Earth (the center of our Galaxy), Eq. (44) gives
$$
{F_{lost} \over F_{expected}} = 0.0116 = 1.16 \%
$$
while, for the same source at a distance of $r_E = 30$ Kpc (the other edge of 
our Galaxy), Eq. (44) gives
$$
{F_{lost} \over F_{expected}} = 0.1044 = 10.44 \%
$$

\section{Discussion and conclusions}

In the present paper we discuss the resonant interaction of a sinusoidal plane polarized 
gravitational wave with the interstellar matter, in the presence of a uniform 
and static in time magnetic field. 
This sort of interaction arises in the parallel and quasiparallel directions 
of propagation of the gravitational wave with respect to the dynamic lines of the 
magnetic field. In any of these cases of propagation and under certain {\em resonant 
conditions}, the trigonometric term in the corresponding 
Hamiltonian function becomes stationary. The dynamical system is therefore trapped 
in a resonance (Kleidis et al 1995). 

Under an appropriate transformation of the affine parameter, which guarantess 
that the particles will continue to move along geodesics, we may show that this 
non-linear interaction problem corresponds to a {\em parametric resonance}. 
In this case the external action (the gravitational 
wave) amounts to a time variation of the frequency parameter of the initial system 
(charged particles in a magnetic field), so that in the overall system the equilibrium 
at rest ($x^1 = 0$) is {\em unstable}. Any deviation from this state, 
however small, is sufficient to lead to a rapidly increasing displacement along 
the peprendicular direction (Landau and Lifshitz 1976). Further analysis shows 
that, in any of the above propagation cases, the overall result would be an 
exponential increase of the perpendicular energy of the particles involved in 
this interaction, as a function of the affine parameter and, hence, of their 
proper-time as well.

However, in this case, except from the primary resonance, there are also many 
secondary resonances, each one associated to a non-zero {\em resonant width} 
(Landau and Lifshitz 1976). This width is given in terms of the dimensionless 
amplitude of the gravitational wave, $ \al$, and the dimensionless ratio, $ \bt 
\simeq \rhv / \om^2$, of the energy density of the interstellar plasma 
over the energy density of the magnetic field (since $ \om^2 \sim E_{magn}$). We 
may also calculate the {\em total measure of the resonant widths}, which is 
related to the total probability for our dynamical system to be trapped in a 
resonance. This is a positive quantity, directly proportional 
to the dimensionless amplitude of the gravitational wave.

Numerical results, in the quasi-parallel case, indicate that an overall 
increase in the perpendicular energy, $I_1$, is also observed in any of the secondary 
resonances. In fact, in some cases this is true even when the value of $I_0$, 
in terms of which we express each resonant 
condition, lies outside of the corresponding resonant width. This 
indicates that the efficiency of the resonant interaction 
mechanism between charged particles and a gravitational wave is considerably larger 
than the one obtained in Papers I and II for the following reason. 

As a gravitational wave propagates into a non-empty space, it encounters 
interstellar clouds with different characteristic temperatures, densities and magnetic 
fields. To each cloud corresponds a specific set of $ \bt $, $\nu$ and $\thv$. 
One expects that, for some of these sets, one of the many resonant 
conditions will be satisfied, even approximately, leading to a parametric resonant 
acceleration of the charged particles. 

We may give an estimate of the gravitational energy flow lost, due to parametric 
resonant interaction of the gravitational wave with the interstellar matter, with 
respect to the flow expected at the Earth in the absence of resonant damping 
mechanisms. The numerical applications indicate that, for a binary source at a 
distance of 30 Kpc, the total loss may be up to $10 \%$ of the flow theoretically 
expected. Of course the calculations are somewhat rough and the results refer to 
the idealized situation in which the wave propagates always parallel to the 
direction of the magnetic field. However, they indicate that the resonant interaction 
of a gravitational wave with the interstellar medium could lead to a reevaluation of 
the today expected values of the wave's amplitude in the neighbourhood of the Earth.

Finally, the analytic results obtained in the present paper generalize the 
corresponding ones of 
Papers I and II, since no averaging technique has been used. We note that the 
main result of the averaging procedure was to smooth out an additional secularily 
increasing term and the trigonometric modulation, which arise in the exact expression 
for the perpendicular energy $I_1( \lm) $. 
\\
\\
{\bf Acknowledgements:} The first author (K K) would like to thank the Greek 
State Scholarships Foundation for the financial support during this work. The 
authors would also like to thank Professor S Persides for his critisism and comments 
on the content of this article. This work is partially supported by the scientific 
program PENED 1451 (Greece).

\section*{References}

\begin{itemize}

\item[] Abramowitz M and Stegun I A 1970 {\em Handbook of Mathematical Functions}
(New York: Dover)
\item[] Bell M 1957 Proceedings of the Glasgow Mathematical Association 3, 132 
\item[] Erdelyi A, Magnus W, Oberhettinger F, Tricomi FG  1953 {\em Higher 
Transcedental Functions} (New York: McGraw-Hill)
\item[] Esposito F P 1971 ApJ 165 165
\item[] Farina D, Pozzoli R, Mannella A, Ronzio R 1993 Phys. Fluids B 5 104
\item[] Gradshteyn I S and Ryzhik I M 1965 {\em Table of Integrals, Series and 
Products} (New York: Academic Press)
\item[] Grishchuk L P and Polnarev A G  1980 {\em Gravitational Waves and their 
Interaction with Matter and Fields} in {\em General Relativity 
and Gravitation - One Hundred Years After the Birth of Albert Einstein} ed A Held
(New York: Plenum) 
\item[] Hillas A M 1984 Ann. Rev. A$\&$A 22 425
\item[] Karimabadi H, Akimoto K, Omidi N et al 1990 Phys. Fluids B 2 606
\item[] Kleidis K, Varvoglis H, Papadopoulos D 1993  A$\&$A 275 309
\item[] Kleidis K, Varvoglis H, Papadopoulos D, Esposito F P 1995  A$\&$A 294 313
\item[] Landau L D and Lifshitz E M 1976 {\em Mechanics} (London: Pergamon Press)
\item[] Lichtenberg A J and Lieberman M A 1983 {\em Regular and Stochastic Motion} 
(New York: Springer)
\item[] Macedo P G, Nelson A H 1983 Phys. Rev. D 28 2382
\item[] Macedo P G, Nelson A H 1990 ApJ 362 584
\item[] Menyuk C R, Drobot A T, Papadopoulos K et al 1987 Phys. Rev. Lett. 58 2071
\item[] Misner C W, Thorne K S and Wheeler J A 1973 {\em Gravitation} (San 
Francisco: Freeman)
\item[] Papadopoulos D, Esposito F P 1985 ApJ 282 330
\item[] Papapetrou A 1974 {\em Lectures on General Relativity} 
(Dordrecht: D Reidel Publishing Co)
\item[] Polnarev A G 1972 Sov. Phys. JETP 35 834
\item[] Smarr L 1979 {\em Sources of Gravitational Radiation} (Cambridge: University 
Press)
\item[] Thorne K S 1987 {\em Gravitational Radiation} in 
{\em Three Hundred Years of Gravitation} eds S W Hawking and W Israel (Cambridge: 
University Press)
\item[] Thorne K S 1995 {\em Gravitational Waves from Compact Bodies} in {\em 
Proceedings 
of IAU Symposium 165, Compact Stars in Binaries} (North Holland: Kluwer Academic 
Publishers)
\item[] Varvoglis H, Papadopoulos D 1992 $ A \& A $ 261 664

\end{itemize}

\end{document}